\documentclass[a4,14pt]{article}
\usepackage[margin=1in]{geometry}
\usepackage{subcaption}
\usepackage{amsmath}
\usepackage{hyperref}
\usepackage{amssymb}
\usepackage{braket}
\usepackage{upgreek}
\usepackage{graphicx}
\usepackage{float}
\usepackage{cite}
\usepackage{esvect}
\usepackage{xcolor}
\begin{document}
\pagenumbering{arabic}
\title{Testing Modified Teleparallel Gravity by Neutrino Oscillations}
\author{H. Yazdani Ahmadabadi\footnote{hossein.yazdani@ut.ac.ir} and H. Mohseni Sadjadi\footnote{mohsenisad@ut.ac.ir}
	\\ {\small Department of Physics, University of Tehran,}
	\\ {\small P. O. B. 14395-547, Tehran 14399-55961, Iran}}
\maketitle
\begin{abstract}
In the framework of modified teleparallel gravity, we study the neutrino flavor conversion.
We show how the presence of spacetime torsion modifies both neutrino flavor oscillation and the matter-enhanced Mikheyev-Smirnov-Wolfenstein (MSW) resonance.
Using data from solar neutrino experiments, we perform an analysis to place new bounds on the modified teleparallel model parameters and neutrino-torsion coupling.
Our numerical results indicate that the resulting modifications to the neutrino survival probability may explain potential observational discrepancies in current and future observatories without invoking new physics or non-standard interactions.
\end{abstract}

\section{Introduction}\label{Sec1}
The discovery of neutrino flavor conversion—first hypothesized to explain the solar neutrino deficit \cite{Pontecorvo,SolarDeficit1,SolarDeficit2} and confirmed by Super-Kamiokande \cite{Super-Kamiokande} and SNO \cite{SNO}—constitutes the definitive proof of non-zero neutrino mass.
In the standard three-flavor picture, oscillations proceed in vacuum while the MSW effect \cite{MSW1,MSW2} governs resonant enhancement in matter \cite{Curved1,Curved.Fornengo,Curved2.Visinelli,Curved3,Curved4,Curved5,Curved6}.
Nevertheless, mild tensions persist between the oscillation parameters inferred from solar and terrestrial datasets, most notably a $\sim2\sigma$ difference in the preferred value of $\Delta m^2_{21}$ \cite{Smirnov,Wurm}.
Such discrepancies motivate explorations of new physics, though conventional extensions often rely on CPT/Lorentz violation \cite{CPTViolat,LoViolat} or sterile neutrinos \cite{Sterile1,Sterile2}, both of which are subject to strong phenomenological bounds.

To address these anomalies, researchers are also exploring interactions beyond the Standard Model.
One approach involves scalar-mediated non-standard interactions (NSIs) \cite{Miranda,Dev,Cordero}, which could modify neutrino mass and matter potentials, producing detectable signals for future experiments like DUNE \cite{DUNE}.
In related models, neutrinos with varying masses have been proposed as a candidate to explain dark energy \cite{Curved7.Sadjadi,Fardon,Kaplan,YazdaniAhmadabadi1,YazdaniAhmadabadi2,YazdaniAhmadabadi3,MohseniSadjadi-GBMaVaN}, linking neutrino mass to a scalar field.
Through interactions with dark sectors, the resulting environment-dependent mass alters the MSW effect and introduces damping signatures, suggesting a possible second-order contribution to the solar neutrino deficit \cite{Curved7.Sadjadi,YazdaniAhmadabadi1,YazdaniAhmadabadi2,YazdaniAhmadabadi3}.

Another promising explanation for the solar neutrino deficit may arise from the coupling between neutrinos and torsion within the framework of the teleparallel equivalent of general relativity (TEGR) \cite{Capolupo1,Capolupo2,Panda,Teleparallel-1,Teleparallel-2,Teleparallel-3,Teleparallel-4,Teleparallel-5,Teleparallel-6,Teleparallel-7}.
These findings indicate that spacetime torsion behaves as an effective potential influencing neutrino propagation, thereby modifying their kinematic properties without invoking NSIs.
Unlike standard GR, which uses curvature, TEGR reformulates gravity through torsion, utilizing the torsion scalar $T$ on a Riemann–Cartan spacetime with the Weitzenb$\ddot{\text{o}}$ck connection \cite{Weitzenbock}.
Extensions to GR, motivated by the Universe's accelerated expansion \cite{UniExp1,UniExp2}, have been explored as alternatives to dark energy \cite{ExpObs1,ExpObs2}.
Within this framework, modified teleparallel gravity—especially $f(T)$ theories
like $f(T) = T + \alpha T^k$ \cite{fofT-1,fofT-2,fofT-3,fofT-4,fofT-5,fofT-6,fofT-7,Ruggiero,Xie,Iorio}—introduces 
nonlinear corrections to the torsion scalar, altering gravitational dynamics to explain late-time acceleration and other deviations from GR.
The model reduces to TEGR when $\alpha=0$. Since TEGR is dynamically equivalent to GR at the classical level and this limit 
produces Einstein gravity despite the presence of a non-vanishing Weitzenbock torsion tensor.

Within this framework, the present paper investigates how spacetime torsion affects neutrino oscillations.
General analytical expressions are derived for both the phase shift in neutrino oscillations and the modified mixing parameters in matter.
These modifications impact fundamental vacuum oscillations and the standard MSW resonance mechanism, highlighting torsion as a non-negligible factor in astrophysical settings.
A key advantage of this approach is the existence of an exact analytical solution to the Dirac equation, enabling direct implementation of the developed formalism.

Moreover, neutrino oscillations  may serve as unique and powerful probes for constraining unexplored regions of the parameter space in extended theories of gravity \cite{Curved3,YazdaniAhmadabadi3,deGiorgi,Parker}.
In the context of $f(T)$ gravity, various observational constraints have been placed on the parameter $\alpha$, ranging from limits based on planetary perihelion precession ($|\alpha| \leq 1.8 \times 10^4~m^2$) \cite{Iorio} to more refined bounds from general relativistic analyses ($|\alpha| \leq 1.2 \times 10^2~m^2$) \cite{SolSysTest1} and increasingly precise methods ($|\alpha| \leq 2.3 \times 10~m^2$) \cite{Ruggiero}.

This work goes beyond heuristic approaches by constructing a theoretical framework within $f(T)$ gravity, aiming to derive novel and tighter constraints on model parameters using solar neutrino flavor conversion.
It provides a framework for testing  $f(T)$ gravity model with current and future neutrino observatories such as JUNO \cite{JUNO}, DUNE \cite{DUNE}, and Hyper-Kamiokande \cite{HK1}.
Also, as the torsion modifies neutrino flavor deficit, potential observational discrepancy in neutrino flavor conversion may be explained in our model without provoking new physics or non-standard interactions.
Furthermore, the mass-squared difference derived in this model offers a plausible explanation for the apparent variation between oscillation parameters extracted from solar and terrestrial neutrino measurements.

This paper is organized as follows.
In Section \ref{Sec2}, we review the theoretical framework of $f(T)$ gravity and solve its field equations for spherically symmetric spacetimes under the ``weak-field approximation'', considering the polynomial density profile.
Section \ref{Sec3} presents the exact solution to the Dirac equation with a torsion-neutrino coupling term, demonstrating its effects on the phase of the neutrino wavefunction.
This section also examines the effects of this new coupling on the neutrino mass and energy terms, which in turn modifies the neutrino Hamiltonian, leading to effective mass and mixing parameters.
In Section \ref{Sec4}, we present the numerical results, compare them against observational data, and provide the ensuing constraints on the model's parameters (e.g., $\alpha$).
We illustrate numerically how different $\alpha$ and couplings result in different survival probabilities.  Concluding remarks are provided in Section \ref{Sec5}.

Throughout the paper, we use units $\hbar=c=1$ and metric signature $(+,-,-,-)$.

\section{Modified Gravity and Spacetime Torsion}\label{Sec2}
Teleparallel gravity reformulates general relativity using torsion instead of curvature, relying on the Weitzenb$\ddot{\text{o}}$ck connection ${\Gamma^\lambda}_{\mu\nu} = e^\lambda_{\hat{a}} \partial_\nu e^{\hat{a}}_\mu$, which has zero curvature and non-zero torsion, so that gravitational effects stem solely from torsional degrees of freedom.
The fundamental dynamical variables in teleparallel gravity are the tetrad fields $e^{\hat{a}}_\mu$, which define an orthonormal basis on the tangent space at each spacetime point \cite{Maluf}.
The corresponding spacetime metric $\text{g}_{\mu\nu}$ is constructed from the tetrads according to
\begin{eqnarray}\label{eqn1}
\text{g}_{\mu\nu} = e^{\hat{a}}_\mu e^{\hat{b}}_\nu \eta_{\hat{a}\hat{b}},
\end{eqnarray}
where Greek indices label spacetime coordinates and Latin indices with a hat correspond to locally inertial Lorentz frames, with $\eta_{\hat{a}\hat{b}} = \mathrm{diag}(+1,-1,-1,-1)$.
The associated torsion tensor is defined as
\begin{eqnarray}\label{eqn2}
\begin{split}
{T^\lambda}_{\mu\nu} &= {\Gamma^\lambda}_{\nu\mu} - {\Gamma^\lambda}_{\mu\nu} \
\\& = e^\lambda_{\hat{a}} \left( \partial_\mu e^{\hat{a}}_\nu - \partial_\nu e^{\hat{a}}_\mu \right).
\end{split}
\end{eqnarray}
The torsion tensor captures the antisymmetric part of the connection and fully encodes the gravitational degrees of freedom in teleparallel gravity.

The dynamics of TEGR are governed by the torsion scalar $T$, constructed from specific quadratic contractions of the torsion tensor
\begin{eqnarray}\label{eqn3}
T = \frac{1}{4} T^{\mu\nu\lambda} T_{\mu\nu\lambda} + \frac{1}{2} T^{\mu\nu\lambda} T_{\nu\mu\lambda} - T^{\mu} T_\mu .
\end{eqnarray}
Under the Lorentz group, the torsion tensor admits an irreducible decomposition into vector, axial, and purely tensorial components \cite{Teleparallel-2}.
From the torsion tensor, one defines the torsion vector
\begin{eqnarray}\label{eqn4}
T_\mu = {T^\lambda}_{\lambda\mu},
\end{eqnarray}
which corresponds to its trace and might play an important role in the teleparallel action.
Furthermore, the axial torsion vector is defined through the dual of the torsion tensor \cite{Shapiro}
\begin{eqnarray}\label{eqn5}
A^\mu = \epsilon^{\alpha\beta\gamma\mu} T_{\alpha\beta\gamma} .
\end{eqnarray}
This quantity denotes the totally antisymmetric torsion component, coupled naturally to fermionic axial currents in curved spacetime.
Its definition employs the contravariant Levi-Civita tensor, $\epsilon^{\alpha\beta\gamma\mu}$, constructed from the weight $+1$ Levi-Civita symbol $\delta^{\alpha\beta\gamma\mu}$ via $\epsilon^{\alpha\beta\gamma\mu} = \frac{\delta^{\alpha\beta\gamma\mu}}{\sqrt{-\text{g}}}$ to guarantee that $\epsilon^{\alpha\beta\gamma\mu}$ transforms as a genuine tensor under general coordinate transformations \cite{Shapiro}.
A widely studied extension of TEGR is obtained by replacing the torsion scalar $T$ with a general function $f(T)$,
in close analogy to $f(R)$ gravity \cite{Capozziello,Sotiriou}.

In Einstein-Cartan theory, one may start from a connection containing both curvature and torsion then
${T^\lambda}_{\mu \nu}\to 0$ naturally recovers GR. In teleparallel model of gravity, and $f(T)$, we start directly from the Weitzenbock connection
where curvature is identically zero. Hence there is no continuous torsion $\to$ GR limit. The gravity itself is encoded in torsion,
therefore turning off torsion removes the gravitational field leaving a Minkowski spacetime rather than a curved GR solution.
The GR limit of the theory is instead obtained through the TEGR limit  $f(T)=T$ (i.e., for $\alpha\to 0$).

The action for $f(T)$ gravity is given by
\begin{eqnarray}\label{eqn6}
\mathcal{S} = \frac{M_p^2}{2} \int d^4x  \left[\mathcal{E}  f(T)\right] + \mathcal{S}_m,
\end{eqnarray}
where $M_p = 1/\sqrt{8\pi G}$ is the reduced Planck mass, $\mathcal{E} = \sqrt{-\det(\text{g}_{\mu\nu})}$ denotes the determinant of the tetrad, and $\mathcal{S}_m$ represents the action of the matter fields.
In contrast to $f(R)$ theories, the field equations of $f(T)$ gravity remain of second order due to the absence of higher-derivative terms 
in the torsion scalar.

We start from the metric \cite{Ruggiero}
\begin{eqnarray}\label{eqn7}
ds^2=e^{A(r)}dt^2-e^{B(r)}dr^2-r^2 d\Omega^2,
\end{eqnarray}
looking for spherically symmetric solutions of the field equations, where $d\Omega^{2}= d\theta^2+ \sin^2 \theta d\phi^2$.
The simplest tetrad yielding the above metric is the diagonal one; however, it does not properly parallelize the static spherically
symmetric geometry in the context of $f(T)$ gravity \cite{Ruggiero,Tamanini}.
So, we will work with a specific tetrad, i.e., the ``good'' tetrad, of the form
\begin{eqnarray}\label{eqn8}
e^{{\hat{a}}}_\mu =
\begin{pmatrix}
	e^{\frac{A}{2}}         &   0                                             &   0                                            &    0         \\
0                 &e^{\frac{B}{2}} \sin{\theta}\cos{\phi}   & -r \cos{\theta}\cos{\phi}  &  r \sin{\theta}\sin{\phi}  \\
0                 &e^{\frac{B}{2}} \sin{\theta}\sin{\phi}  & -r  \cos{\theta}\sin{\phi}&  -r  \sin{\theta}\cos{\phi} \\
0                 &e^{\frac{B}{2}} \cos{\theta}  & r \sin{\theta}&  0\\
\end{pmatrix},
\end{eqnarray}
that does not yield a constant torsion scalar. This tetrad eliminates inertial contributions and yields a consistent spherically
symmetric solution in $f(T)$ gravity \cite{Tamanini}. In this geometry, the axial torsion vector vanishes identically,
so axial couplings do not contribute.
According to this tetrad field, the torsion scalar is explicitly given by
\begin{eqnarray}\label{eqn9}
T(r) =\frac{2e^{-B(r)}(1+e^{B(r)/2})}{r^{2}}\left[1+e^{B(r)/2}+r A'(r)\right].
\end{eqnarray}

Variation of the action with respect to the tetrad $e^{\hat{a}}_\mu$ gives the following set of equations \cite{Ruggiero}:
\begin{eqnarray}\label{eqn10}
\mathcal{E}^{-1}\partial_\mu(\mathcal{E}~e_{\hat{a}}^{\rho} {S_{\rho}}^{\mu\nu})f_T + e_{{\hat{a}}}^{\lambda} {S_{\rho}}^{\nu\mu} {T^{\rho}}_{\mu\lambda} f_T
+  e_{\hat{a}}^{\rho}  {S_{\rho}}^{\mu\nu}(\partial_\mu T) f_{TT} + \frac{1}{4} e_{\hat{a}}^{\nu} f = 4\pi G e_{\hat{a}}^{\mu} {\mathcal{T}}_\mu^\nu,
\end{eqnarray}
where ${\cal{T}}^\nu_\mu$ is the matter energy-momentum tensor and subscript $T$ denotes differentiation with respect to torsion scalar $T$.
In the rest of the paper, we consider $f(T) = T + \alpha T^k$ \cite{Ruggiero,Xie,Iorio}, where $\alpha$ is a constant parameter, implying the deviation of such theories from the standard TEGR.
Indeed, for actual physical situations such as in the solar system, the gravity is relatively weak compared to the strong-field regimes near black holes or neutron stars.
So, we expand the metric exponentials in the weak-field limit to linear order and write
\begin{eqnarray}\label{eqn11}
e^{A(r)} \approx 1 + A(r), && e^{B(r)} \approx 1 + B(r).
\end{eqnarray}
By substituting the tetrads from Eq. (\ref{eqn8}) into the corresponding field equations and linearizing them with respect to the functions $A(r)$ and $B(r)$, we derive a simplified set of equations
\begin{eqnarray}\label{eqn12}
r^3 B'(r) + r^2 B(r) + 96 \alpha - 8\pi G r^4 \rho=0,
\end{eqnarray}
\begin{eqnarray}\label{eqn13}
r^3 A'(r) + r^2 B(r) - 32\alpha + 4 \pi G r^4 p =0,
\end{eqnarray}
and
\begin{eqnarray}\label{eqn14}
r^4 A''(r) - r^3 \left[A'(r) + B'(r)\right] + 2r^2 B(r) -128 \alpha =0,
\end{eqnarray}
where $\rho$ and $p$ are the energy density and pressure of matter component, and prime denotes differentiation with respect to the radial coordinate $r$.
Throughout the present analysis, we set the power of the second term in the $f(T)$ function to $k=2$, i.e., we investigate $f(T) = T + \alpha T^2$.
The simplified equations are then solved under both vacuum conditions and the polynomial matter density profile, with the resulting solutions shown below.

\subsection{Vacuum Solution}\label{Subsec2.1}
In vacuum, exact solutions of Eqs. (\ref{eqn12})-(\ref{eqn14}) are given by
\begin{eqnarray}\label{eqn15}
\begin{split}
&A(r)= -\frac{32 \alpha}{r^2} - \frac{C}{r}, \\&
B(r)= \frac{96 \alpha}{r^2} + \frac{C}{r},
\end{split}
\end{eqnarray}
where $C$ is an integration constant, and we have considered $\rho = p = 0$.
Setting $C = 2GM$ gives the weak-field limit of the Schwarzschild geometry supplemented by corrections generated by the non-linear torsion term $\alpha T^2$.
The $r^{-2}$ contributions appearing in (\ref{eqn15}) arise directly from solving Eqs. (\ref{eqn12})-(\ref{eqn14}) in the presence of $\alpha T^2$ term.
In the limit  $\alpha\to 0$, these corrections disappear and the standard weak-field Schwarzschild solution is recovered, as expected from equivalence between TEGR and GR at the classical level. 
We note that the torsion scalar $T$ and the Ricci scalar $R$ differ only by a total derivative.
Consequently, the TEGR action differs from the Einstein-Hilbert action only by a boundary term, and under appropriate boundary conditions both theories yield exactly the same gravitational field equations.

Therefore, we have
\begin{eqnarray}\label{eqn16}
ds^2 = \left(1- \frac{2GM}{r} - \frac{32\alpha}{r^2}\right)dt^2 - \left(1+ \frac{2GM}{r} + \frac{96\alpha}{r^2}\right)dr^2 - r^2 d\Omega^2.
\end{eqnarray}
Plugging the components (\ref{eqn15}) into the general expression for the tetrad (\ref{eqn8}), we find the torsion scalar
\begin{eqnarray}\label{eqn17}
T(r) = \frac{8}{r^2} -\frac{128 \alpha}{r^4},
\end{eqnarray}
where the last term is due to the higher-order torsion (proportional to $\alpha T^2$) effects.

Eq.(\ref{eqn17}) shows that the torsion scalar remains nonzero in the vacuum. This behavior is characteristic of teleparallel gravity, where torsion represents the gravitational field itself.
This should be contrasted with Einstein-Cartan theory, in which torsion is algebraically sourced by spin density of matter and therefore typically vanishes outside matter distribution. 

Working in the equatorial plane by setting $\theta = \pi/2$ and $\phi = 0$ simplifies the radial distance to $x = r$.
This simplification means that $T_1$ below depends solely on the radial coordinate, which will facilitate the derivation of subsequent relations for neutrino propagation.
Thus, the only non-zero component of the torsion vector is obtained as
\begin{eqnarray}\label{eqn18}
\begin{split}
&T_1(r) \equiv {T^\lambda}_{\lambda 1} \\& = -\frac{4}{r} - \frac{3GM}{r^2} - \frac{128\alpha}{r^3}.
\end{split}
\end{eqnarray}

\subsection{Solution inside Matter}\label{Subsec2.2}
The present study adopts a static, spherically symmetric, non-relativistic matter distribution with radial density $\rho(r)$, neglecting pressure contributions in the weak-field regime.
Although realistic solar models require helioseismic data and stellar structure theory \cite{Bahcall}, simplified analytic profiles are frequently employed for theoretical work \cite{NASAWebsite}, given here by a polynomial density model
\begin{eqnarray}\label{eqn19}
	\rho(r) =
	\begin{cases}
		\rho_0 \left(a \left(\frac{r}{R_\odot}\right)^4 + b \left(\frac{r}{R_\odot}\right)^3 + c \left(\frac{r}{R_\odot}\right)^2 + d \left(\frac{r}{R_\odot}\right) + e\right) & (\text{for}~r\leq R_\odot) \\
		0 & (\text{for}~r > R_\odot)
	\end{cases}
	,
\end{eqnarray}
where $a$, $b$, $c$, $d$, and $e$ are dimensionless fitting parameters \cite{NASAWebsite}, $\rho_0$ sets the overall density scale, and $R_\odot$ denotes the solar radius.
Introducing the dimensionless radial coordinate $R = r/R_\odot$, the interior of the Sun corresponds to the interval $R \in [0,1]$.
Solving the modified gravitational field equations (\ref{eqn12})-(\ref{eqn14}) gives
\begin{eqnarray}\label{eqn20}
	\begin{split}
		&A(R)= -\frac{32 \alpha}{R^2 R_\odot^2} - \frac{2GM}{R R_\odot} + 2\pi G \rho_0 R_\odot^2 \left(\frac{2}{21} a R^6 - \frac{2}{15} b R^5 + \frac{1}{5} c R^4 - \frac{1}{3} d R^3\right), \\&
		B(R)= \frac{96 \alpha}{R^2 R_\odot^2} + \frac{2GM}{R R_\odot} + 2\pi G\rho_0 R_\odot^2 \left(\frac{4}{7} a R^6 - \frac{2}{3} b R^5 + \frac{4}{5} c R^4 - d R^3 + \frac{4}{3} e R^2\right).
	\end{split}
\end{eqnarray}
These expressions show that, in the weak-field regime, the Schwarzschild solution is modified by additional terms proportional to the $f(T)$ parameter $\alpha$.
In the vacuum limit, obtained by taking $\rho_0 \to 0$, the solution reduces to the case given in Eq.(\ref{eqn15}).
Consequently, the $\alpha$-dependent corrections decay at large distances, ensuring asymptotic convergence to the Schwarzschild solution.

The polynomial constants modify both the timelike and radial components of the metric in a radius-dependent manner.
From the metric solutions (\ref{eqn20}), we can derive important geometrical quantities.
As a consequence, the above results can be applied to obtain the torsion scalar
\begin{eqnarray}\label{eqn21}
T(R) = \frac{8}{R^2 R_\odot^2} - \frac{128 \alpha}{R^4 R_\odot^4} -\frac{32}{3} \pi G \rho_0 e.
\end{eqnarray}
Remarkably, except for a constant shift proportional to the parameter $e$, this expression is independent of the detailed matter density profile $\rho(r)$, depending explicitly only on the torsion parameter $\alpha$.
This indicates that, in this case, torsion effects are dominated by the purely geometric $T^2$ contributions, while the coupling to matter, only through the global constant term $\rho_0 e$, plays only a subleading and non-dynamical role.
Furthermore, the non-zero component of the torsion vector takes a more complicated form
\begin{eqnarray}\label{eqn22}
T_1(R)= -\frac{4}{R R_\odot} -\frac{3GM}{R^2 R_\odot^2} - \frac{128 \alpha }{R^3 R_\odot^3} -  \pi G\rho_0 R_\odot \left(\frac{12}{7} a R^5 - 2b R^4 + \frac{12}{5} c R^3 - 3d R^2\right).
\end{eqnarray}
For the above solutions to remain physical, $\alpha$ must be small enough that the metric components do not exhibit unphysical singularities or sign changes outside the horizon.
Solar system tests \cite{SolSysTest1,SolSysTest2,SolSysTest3,SolSysTest4,SolSysTest5} can place bounds on $\alpha$.
In the present work, however, we try to put bounds on this parameter using the neutrino oscillation observations \cite{Borexino}.

\section{Torsion Effects on Neutrino Oscillations}\label{Sec3}
This study investigates the modification of neutrino flavor oscillations by spacetime torsion within the framework of teleparallel gravity.
We note that the gravitational field can also affect neutrino oscillations through the bending of neutrino trajectories \cite{ref1,ref2,ref3}.
The effect studied here is different and arises from the direct coupling of torsion to the neutrino spinor field, modifying the neutrino effective mass and energy.

The torsion tensor ${T^{\rho}}_{\mu\nu}$ admits a standard irreducible decomposition into a vector part $T_\mu = {T^\nu}_{\nu\mu}$ (Eq.~(\ref{eqn4})), an axial-vector part $A^\mu = \frac{1}{6}\epsilon^{\mu\nu\rho\sigma}T_{\nu\rho\sigma}$ (Eq.~(\ref{eqn5})), and a purely tensorial part $t_{\lambda\mu\nu}$ \cite{Teleparallel-2,Capolupo1,Capolupo2,Panda}.
This decomposition yields an effective torsion-induced Lagrangian of the form $\mathcal{L}^{\text{VA}} \propto \kappa^V T_\mu\, \bar\psi\gamma^\mu\psi + \kappa^A A_\mu\, \bar\psi\gamma^\mu\gamma^5\psi$, where $\kappa^{V,A}$ denote the neutrino-torsion coupling coefficients.
For Dirac neutrinos, the non-vanishing vector current $\bar\nu_D\gamma^\mu\nu_D$ couples to $T_\mu$, leading to a modified Dirac equation $\big( i\gamma^\mu D_\mu - m + \kappa^V \gamma^\mu T_\mu \big)\nu_D = 0$, which describes propagation in an external vector potential.
Conversely, for Majorana neutrinos the vector current $\bar\nu_M\gamma^\mu\nu_M$ vanishes identically \cite{Akhmedov}, leaving only a possible axial-vector coupling $\bar\nu_M\gamma^\mu\gamma^5\nu_M$ to $A_\mu$.
The covariant derivative $D_\mu = \partial_\mu + \Gamma_\mu$ incorporates the spin connection $\Gamma_\mu = \tfrac{1}{8} [\gamma^{\hat{a}},\gamma^{\hat{b}}] e^\nu_{\hat{a}} e_{\hat{b} \nu ; \mu}$, whose contraction $\gamma^{\hat{a}} e^\mu_{\hat{a}} \Gamma_\mu$ is proportional to $A_\mu$ \cite{Curved1,Curved6}.
However, in the simplest spherically symmetric teleparallel geometries—such as TEGR with the well-behaved tetrad of Eq.~(\ref{eqn8})—the axial-vector component vanishes ($A^\mu = 0$) while $T_\mu$ remains non-zero \cite{Piriz}.
Consequently, the spin-connection term does not contribute, and the only surviving torsion coupling in this geometry is the vector interaction.
Since the vector current vanishes for Majorana neutrinos, no torsion-induced modification occurs for Majorana fields in this setting; the subsequent analysis is therefore restricted to Dirac neutrinos.

Following the above considerations, we model torsion effects from an astrophysical source through a vector field $T_\mu$.
The neutrino propagation is then described by the action
\begin{eqnarray}\label{eqn23}
\begin{split}
\mathcal{S}_\nu &= \int d^4 x~e \bigg[- \sum_{\alpha,\beta} \upnu^{\dagger}_{\alpha L} m_{\alpha \beta} \upnu_{\beta R} - \sum_{\alpha,\beta} \upnu^{\dagger}_{\alpha R} m^{*}_{\beta \alpha} \upnu_{\beta L} \\& + i\sum_\alpha
\left(\upnu_{\alpha L}^{\dagger} \sigma^\mu_L D_\mu \upnu_{\alpha L}  + \upnu_{\alpha R}^{\dagger} \sigma^\mu_R D_\mu \upnu_{\alpha R} \right) \\& +
\sum_{\alpha,\beta} T_\mu  \left(\kappa_{\alpha\beta}~\upnu_{\alpha L}^{\dagger} \sigma^\mu_L \upnu_{\beta L}  + \kappa^*_{\beta \alpha}~\upnu_{\alpha R}^{\dagger} \sigma^\mu_R \upnu_{\beta R} \right)\bigg].
\end{split}
\end{eqnarray}
The mass terms are defined by the components $m_{\alpha\beta}$ of a $3 \times 3$ flavor-basis matrix acting on the left- and right-handed two-component spinors, $\upnu_{\alpha (L,R)}$.
The chiral structure is encoded in the right- and left-handed Pauli matrices, $\sigma^\mu_R= \left(\sigma^0,\sigma^1,\sigma^2,\sigma^3\right)$ and $\sigma^\mu_L= \left(\sigma^0,-\sigma^1,-\sigma^2,-\sigma^3\right)$, respectively.
Furthermore, the Lorentz vector field for spinors is expressed as $\bar{\psi} \gamma^\mu \psi = \psi^\dagger_L \sigma^\mu_L \psi_L + \psi^\dagger_R \sigma^\mu_R \psi_R$ \cite{Cott-Green}, as reflected in the action's third line.
The neutrino-torsion couplings are taken to be the elements $\kappa_{\alpha\beta}$ of a flavor-space coupling matrix, which is assumed to be diagonal throughout this work.
For notational simplicity, the superscript $V$ on the coupling coefficient has been dropped throughout.

\subsection{Vacuum Neutrino Oscillation}\label{Subsec3.1}
Neutrino oscillation—the detection of a flavor distinct from that produced at the source—arises from the mixing of mass and flavor eigenstates, expressed as
\begin{eqnarray}\label{eqn24}
\upnu_{\alpha (L,R)} = \sum_{i} U^{* (L,R)}_{\alpha i} \upnu_{i (L,R)},
\end{eqnarray}
where $U_{\alpha i}$'s are components of the Pontecorvo-Maki-Nakagawa-Sakata (PMNS) unitary mixing matrix.
Massless neutrinos would have plane-wave solutions of negative helicity.
However, by introducing the masses for neutrinos, we may redefine its wavefunctions as
\begin{eqnarray}\label{eqn25}
\begin{split}
&\upnu_{\alpha L}(x,t) = e^{-iE_\nu(t-t_0)} e^{+iE_\nu(x-x_0)} h_{\alpha}(x)
	\begin{pmatrix}
		0\\
		1\\
	\end{pmatrix},
	\\&
\upnu_{\alpha R}(x,t) = e^{-iE_\nu(t-t_0)} e^{+iE_\nu(x-x_0)} g_{\alpha}(x)
	\begin{pmatrix}
		0\\
		1\\
	\end{pmatrix}.
\end{split}
\end{eqnarray}
By varying the neutrino action (\ref{eqn23}) with respect to the left- and right-handed spinors, and using relations (\ref{eqn25}), the equations of motion are given by
\begin{eqnarray}\label{eqn26}
i\frac{d h_\alpha(x)}{dx} - m_{\alpha \beta}~g_\beta(x) + \kappa_{\alpha\beta} T_1(x) h_{\beta}(x) = 0,
\end{eqnarray}
and
\begin{eqnarray}\label{eqn27}
2E_\nu g_\alpha(x) - i \frac{d g_\alpha(x)}{dx} - \kappa^*_{\beta\alpha} T_1(x) g_\beta(x)  - m^*_{\beta\alpha} h_\beta(x) = 0,
\end{eqnarray}
where we have adopted the Einstein summation convention; summation signs are restored solely when needed.
Since the magnitude of the second term (proportional to the mass term) is much lower than the first one (which includes neutrino energy) in (\ref{eqn27}), we can neglect that (i.e., $|dg_\alpha/dx| \ll 2E_\nu g_\alpha$).
Hence, under the assumption of a diagonal neutrino-torsion coupling matrix, substituting the expression for $g_\alpha(x)$ obtained from Eq.~(\ref{eqn27}) into Eq.~(\ref{eqn26}) yields directly the equation of motion
\begin{eqnarray}\label{eqn28}
i\frac{d h_i(x)}{dx} - \frac{\hat{m}^2_i}{2\hat{E}_\nu} h_i(x) = 0,
\end{eqnarray}
where we have used the characteristics $m_{\alpha\beta} = U^*_{\alpha i} U_{\beta i} m_i$, $\kappa_{\alpha\beta} = U^*_{\alpha i} U_{\beta i} \kappa_i$, and $h_\alpha(x) = U^*_{\alpha i} h_i(x)$ (based on the mixing (\ref{eqn24})).

Within the framework of $f(T)$ gravity, the presence of spacetime torsion generates additional terms in the equation of motion that couples to the neutrino's spinor field.
These terms are mathematically equivalent to a shift in both the neutrino energy and its effective mass-squared, motivating the definition of the effective quantities $\hat{E}_\nu$ and $\hat{m}_i^2$.
Applying the unitarity condition of the PMNS mixing matrix, i.e., $\sum_{\alpha} |U_{\alpha i}|^2 =1$ for a fixed mass eigenstate index $i$, the redefinition of the neutrino energy, i.e.,
\begin{eqnarray}\label{eqn29}
\hat{E}^{(i)}_\nu \equiv  E_\nu - \frac{1}{2} \sum_{\alpha} |U_{\alpha i}|^2 \kappa_i T_1(x) = E_\nu - \frac{1}{2} \kappa_i T_1(x),
\end{eqnarray}
can be interpreted as an effective potential induced by spacetime torsion.
We emphasize that $E_\nu$ denotes the conserved energy associated with the timelike Killing vector field of the static spacetime.
The effective energy $\hat{E}_\nu(x)$ arises from recasting the Dirac equation into a Schr$\ddot{\text{o}}$dinger-like form for the spatial evolution of the wavefunction.
The spatial dependence of $\hat{E}_\nu$ does not imply a violation of energy conservation; rather, it reflects the exchange of gravitational/torsional potential energy between the neutrino field and the background geometry, analogous to the role of the matter potential $V(x)$ in the standard MSW effect.
Since oscillation probabilities depend solely on the oscillation phase difference, this redefinition is phenomenologically sound and ensures the correct covariant calculation of flavor transitions.

Similarly, the effective mass-squared acquires torsion-induced corrections proportional to both the neutrino energy and the coupling parameter, namely
\begin{eqnarray}\label{eqn30}
\begin{split}
\hat{m}_i^2 & \equiv m_i^2 - 2\hat{E}^{(i)}_\nu \sum_{\alpha} |U_{\alpha i}|^2 \kappa_i T_1(x) \\& = m_i^2 - 2E_\nu  \kappa_i T_1(x) + \mathcal{O}\left({\kappa_i}^2 T_1^2(x)\right),
\end{split}
\end{eqnarray}
where we have substituted the definition of $\hat{E}_\nu$ from Eq.~\eqref{eqn29} and retained terms up to linear order in the coupling $\kappa_i$.
We have used again the unitarity condition $\sum_{\alpha} |U_{\alpha i}|^2 =1$.
Consequently, torsion may induce novel oscillation patterns or phase shifts even in the absence of ordinary matter, offering a purely geometric modification to the standard oscillation framework.
At leading order, this correction introduces an energy-dependent shift to the effective neutrino mass spectrum, which directly alters the accumulated oscillation phase.
This energy dependence is particularly noteworthy, as it deviates from the standard vacuum paradigm where masses are strictly constant parameters.
Such modifications could, in principle, provide an alternative interpretation of anomalous spectral features observed in low-energy solar or reactor neutrino data, though a detailed phenomenological analysis lies beyond the scope of the present work.

Now, we solve the equation (\ref{eqn28}) easily to obtain the neutrino's wavefunction
\begin{eqnarray}\label{eqn31}
\upnu_{i}(x,t) = e^{-iE_\nu\left[(t-t_0)-(x-x_0)\right]} e^{-i \Phi_i(x)} \upnu_{i}(x_0,t_0).
\end{eqnarray}
Therefore, the quantity
\begin{eqnarray}\label{eqn32}
\Phi_i(x) = \int^x_{x_0} \left[\frac{m_i^2}{2E_\nu - \kappa_i T_1(x)} -  \kappa_i  T_1(x)\right] dx
\end{eqnarray}
is defined as the modified phase of oscillations in vacuum.
This oscillation phase is similar to that flat spacetime (the mass-squared term), but is modified due to the background torsion.

In the specific regime, where $\kappa T_1(x) \ll E_\nu$, the phase difference $\Phi_{ij}$ between neutrino mass eigenstates is given by
\begin{eqnarray}\label{eqn33}
\Phi_{ij} = \frac{\Delta m_{ij}^2 \left(x - x_0\right)}{2E_\nu} - \Delta \kappa_{ij} \int^x_{x_0} T_1(x) dx.
\end{eqnarray}
Here, $\Delta \kappa_{ij} = \kappa_i - \kappa_j$ represents the difference in how each mass eigenstate couples to the spacetime torsion field.
This formulation demonstrates that the oscillation phase acquires a novel contribution from the integral of the torsion field along the neutrino's path.

Another situation arises if the torsion coupling \(\kappa\) is universal, i.e., ($\kappa_i = \kappa_j$).
In this case, the coupling difference vanishes, and the phase difference then simplifies to
\begin{eqnarray}\label{eqn34}
\Phi^{(0)}_{ij} = \frac{\Delta m^2_{ij} (x-x_0)}{2 E_\nu}.
\end{eqnarray}
So, a non-universal coupling to torsion is necessary to have observable effects, as it is the coupling difference $\Delta \kappa_{ij}$ that drives the deviation from standard neutrino oscillations.

Since neutrino mass eigenstates are linear superpositions of flavor eigenstates via the PMNS mixing matrix, any flavor-dependent torsion coupling—analogous to a torsion charge—results in distinct effective couplings $\kappa_i$ for the mass eigenstates, with $\kappa_i \ne \kappa_j$ arising simply because $\nu_i$ and $\nu_j$ contain different proportions of flavors that interact with torsion unequally.
We emphasize that assigning such flavor-dependent couplings formally constitutes a violation of the Weak Equivalence Principle within the torsion sector, although the universal coupling to the metric background remains fully preserved.
This scenario is conceptually analogous to postulating a new fifth force charge for neutrinos, akin to the flavor-dependent weak potential in the MSW effect.

To apply this general result to a physical scenario, such as neutrino propagation from the surface of the Sun, a specific model for the component $T_1(x)$ is required.
By adopting the obtained torsion profile (Eq.(\ref{eqn15})), the phase difference in vacuum can be expressed as
\begin{eqnarray}\label{eqn35}
\begin{split}
\Phi_{ij}(R) &= \frac{\Delta m_{ij}^2 R_\odot \left(R - 1\right)}{2E_\nu} \\& + \Delta \kappa_{ij} \left[\frac{(R - 1) \left(3G M_\odot R R_\odot + 64 \alpha (R + 1)\right)}{R^2 R_\odot^2} + 4 \ln (R)\right].
\end{split}
\end{eqnarray}
The second term includes the geometric and gravitational contributions of the torsion.
From a physical standpoint, these results suggest that spacetime torsion acts as an effective medium through which neutrinos propagate, altering their kinematics without invoking new particle interactions (non-standard interactions).
The torsion-induced terms can be viewed as a gravitationally generated background field that distinguishes between neutrino mass eigenstates through the coupling constants $\kappa_i$.
This provides a mechanism through which modified gravity theories may leave observable imprints in neutrino oscillation experiments.

\subsection{Modified Matter Effects}\label{Subsec3.2}
The MSW effect \cite{MSW1,MSW2} arises from the combined influence of vacuum neutrino oscillations and coherent forward scattering of neutrinos with electrons in a medium, resulting in a resonant enhancement of flavor transitions.
Electron-neutrinos are subject to an additional matter potential $V(x) = \sqrt{2} G_F N_e$, where $G_F$ denotes Fermi's constant and $N_e$ represents the local electron number density.
Upon incorporating the $\nu_e$-$e^-$ weak interaction contribution, the torsion-modified Schr$\ddot{\text{o}}$dinger-like evolution equation (\ref{eqn28}) assumes the following form for propagation within matter:
\begin{eqnarray}\label{eqn36}
i\frac{d h_i(x)}{dx} = \frac{\hat{m}^2_i}{2\hat{E}_\nu} h_i(x)  + V(x) U_{ei} U^*_{ej} h_{j}(x).
\end{eqnarray}
Without loss of generality, we have
\begin{eqnarray}\label{eqn37}
\begin{split}
& i \frac{d}{dr}
	\begin{pmatrix}
		h_1\\
		h_2\\
		h_3
	\end{pmatrix}
	\\& = \frac{1}{2\hat{E}_\nu}
	\begin{pmatrix}
		2\hat{E}_\nu V(r) c_{12}^2 c_{13}^2 & 2\hat{E}_\nu V(r) c_{12} s_{12} c_{13}^2  & 2\hat{E}_\nu V(r) c_{12} c_{13} s_{13}\\
		2\hat{E}_\nu V(r) c_{12} s_{12} c_{13}^2 & 2\hat{E}_\nu V(r) s_{12}^2 c_{13}^2 + \Delta \hat{m}_{21}^{2} &  2\hat{E}_\nu V(r) s_{12} c_{13} s_{13}\\
		2\hat{E}_\nu V(r) c_{12} c_{13} s_{13} &  2\hat{E}_\nu V(r) s_{12} c_{13} s_{13} & 2\hat{E}_\nu V(r) s^2_{13} + \Delta \hat{m}^{2}_{31}
		\end{pmatrix}
	\begin{pmatrix}
		h_1\\
		h_2\\
		h_3
	\end{pmatrix}
,
\end{split}
\end{eqnarray}
where, neglecting the term proportional to $\mathcal{O}\left({\kappa_i}^2 T_1^2\right)$ in (\ref{eqn30}), we have defined $\Delta\hat{m}^{2}_{ij} (r) \simeq \Delta m_{ij}^{2} -2 E_\nu \Delta \kappa_{ij} T_1(r)$ as the torsion-induced mass-squared difference.
As previously stated, we restricted our analysis to directions that lie exactly on the $x$-axis in equatorial plane, where the radial distance is $x=r$.
For such a case, we set $\left(\theta_{\text{polar}},\phi_{\text{azimuthal}}\right)=\left(\pi/2,0\right)$.

The effective mass and mixing parameters for neutrinos inside matter with background torsion are obtained by diagonalizing the effective Hamiltonian.
A common approximation for many experiments is the two-flavor scenario.
For solar neutrinos, the condition $|\Delta \hat{m}^{2}_{31}| \gg 2\hat{E}_\nu V(r)$ holds, allowing flavor transitions to be described by an effective two-neutrino model, with the third eigenstate decoupling from the other two \cite{Agarwalla,Martinez-Mirave}.
Therefore, the effective Hamiltonian governing the propagation of neutrinos through matter is reduced to
\begin{eqnarray}\label{eqn38}
	\mathcal{H}_m = \frac{1}{4\hat{E}_\nu} \left[\Delta \hat{m}^{2}
	\begin{pmatrix}
		-\cos 2\theta&\sin 2\theta \\
		\sin 2\theta & \cos 2\theta
	\end{pmatrix}
	+ 2\hat{E}_\nu V(r)
	\begin{pmatrix}
		1&0\\
		0&-1
	\end{pmatrix}
	\right],
\end{eqnarray}
where $\Delta \hat{m}^{2} \equiv \Delta \hat{m}^{2}_{21}$ for 2$\nu$-case.
Then, the resulting neutrino oscillation parameters, after accounting for the torsion-induced couplings, are
\begin{eqnarray}\label{eqn39}
\Delta m^2_m = \Delta \hat{m}^{2} \sqrt{\sin^2 2\theta + \left[\cos 2 \theta - \frac{2\hat{E}_\nu V(r)}{\Delta \hat{m}^{2}}\right]^2},
\end{eqnarray}
and
\begin{eqnarray}\label{eqn40}
\sin 2\theta_{m} = \frac{\sin 2\theta}{\sqrt{\sin^2 2\theta + \left[\cos 2 \theta - \frac{2\hat{E}_\nu V(r)}{\Delta \hat{m}^{2}}\right]^2}}.
\end{eqnarray}
These key oscillation parameters are naturally influenced by the Sun's electron number density.
Moving forward, it is simpler to treat neutrinos as if they were in a vacuum, but with re-defined mass and mixing parameters $\Delta m^2_m$ and $\theta_{m}$.
It is evident that, even in vacuum, where the electron density $N_e$ approaches zero, the presence of torsion modifies the neutrino
mass-squared splitting, i.e., $\Delta m^2_m \to \Delta \hat{m}^2 \ne \Delta m^2$. For the allowed ranges $|\alpha|\lesssim 5m^2$
the torsion-induced correction to $\Delta m_{31}^2$ is negligible compared to its vacuum value ($\sim 2.5 \times 10^{-3} \text{eV}^2)$,
so the two-flavor approximation remains valid for solar neutrinos.

\section{Numerical Results}\label{Sec4}
In the context of $f(T)$ gravity, spacetime torsion couples directly to neutrino spinor fields, leading to an additional term in the neutrino equation of motion, see Section \ref{Sec3}.
This coupling can be reinterpreted as an effective shift in both the neutrino energy and its mass-squared.
The resulting modifications alter the oscillation phases in a manner analogous to matter-induced effects, but with a purely geometric origin, see Subsection \ref{Subsec3.1}.
Unlike the MSW effect or NSI in matter, the torsion-induced corrections persist in vacuum and reflect properties of spacetime torsion through the higher-order terms in $f(T) = T +\alpha T^k$.
Consequently, neutrino oscillations provide a sensitive probe of teleparallel gravity effects, potentially allowing constraints on torsion couplings from current and future neutrino experiments.

For the numerical calculations, we use the neutrino mass and mixing parameters obtained from the NuFit-6.0 global analysis \cite{Esteban}, which are listed in Table \ref{table1}.
These values incorporate the latest experimental constraints from oscillation data, including reactor, accelerator, and solar neutrino measurements, ensuring consistency with current phenomenological benchmarks.
\begin{table}[H]
	\begin{center}
		\tiny
		\caption{\footnotesize{The global fit of three-flavor oscillation parameters at the best-fit and 1$\sigma$ confidence level, performed by NuFit-6.0 \cite{Esteban}.
		}}
		\label{table1}
		\begin{tabular}{|c|c|c|c|c|c|c|}
			\hline
			\hline
			& $\sin^2\left(\theta_{12}\right)$ & $\Delta m^2_{21} [10^{-5} \text{eV}^2]$ \\
			\hline \hline
			SK+SNO \cite{Nakajima,KamLAND} & $0.306^{+0.014}_{-0.014}$ & $6.11^{+1.21}_{-0.68}$ \\
			\hline
			KamLAND \cite{Nakajima,KamLAND} & $0.316^{+0.034}_{-0.026}$ & $7.54^{+0.19}_{-0.18}$ \\
			\hline
			Combined (NuFit-6.0) \cite{Esteban} & $0.308^{+0.012}_{-0.011}$ & $7.49^{+0.19}_{-0.19}$ \\
			\hline
		\end{tabular}
	\end{center}
\end{table}

\begin{figure}[H]
	\centering
	\includegraphics[scale=0.35]{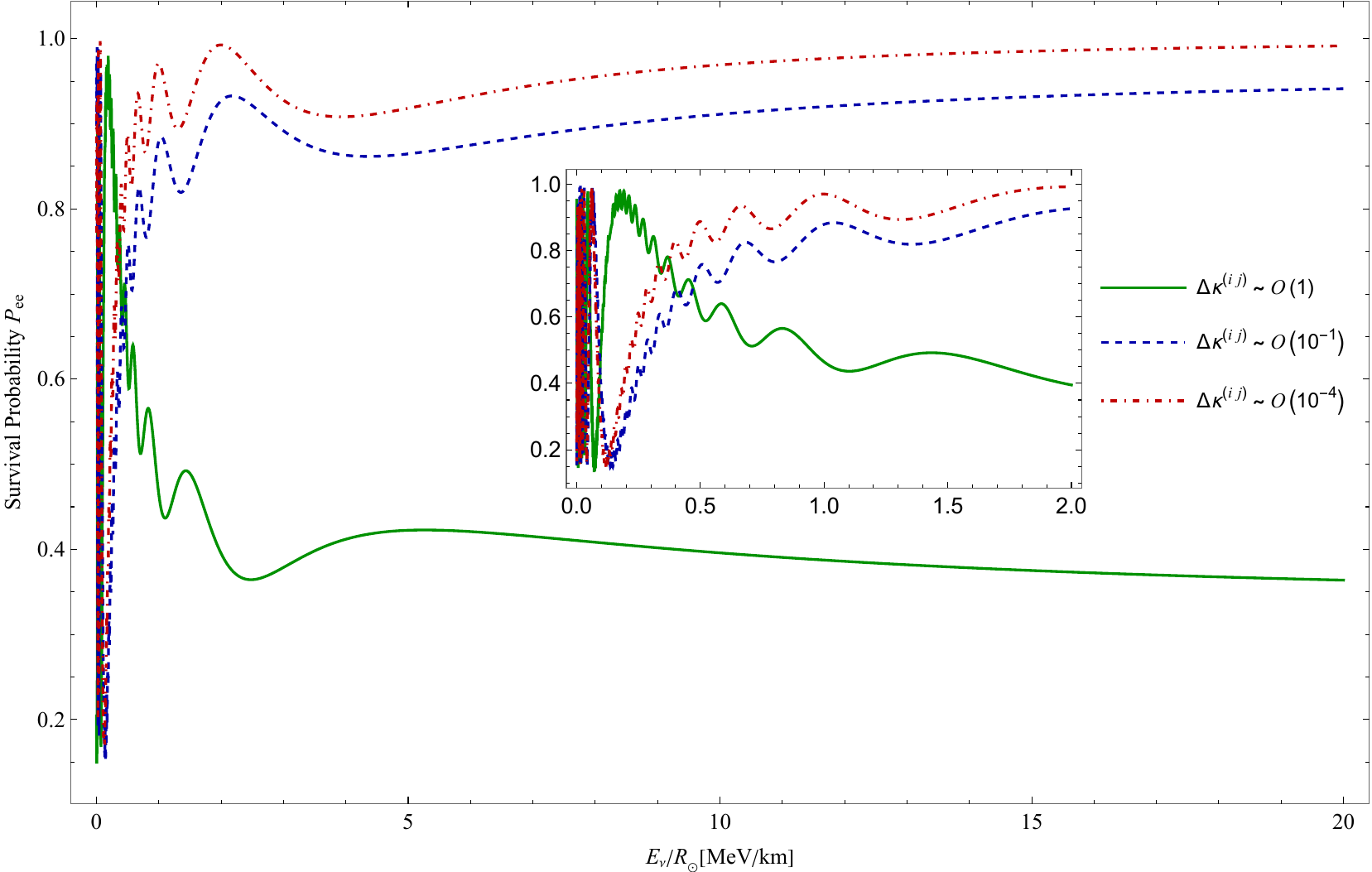}
	\caption{\footnotesize{This figure displays the $\nu_e$ survival probability in vacuum as a function of $E_\nu/R_\odot$ for different $\Delta\kappa_{ij}$ values, with $\alpha$ fixed at $\mathcal{O}(1~m^2)$.}}
	\label{fig1}
\end{figure}

We begin by investigating the influence of various neutrino-torsion coupling strengths on oscillation phenomena in vacuum.
Figure \ref{fig1} demonstrates the qualitative consequences of varying the coupling difference parameter, $\Delta\kappa_{ij}$.
The plot clearly indicates that larger $\Delta\kappa_{ij}$ values result in a suppression of the oscillation amplitude, an effect that becomes more pronounced at higher neutrino energies.

\begin{figure}[H]
	\centering
	\includegraphics[scale=0.45]{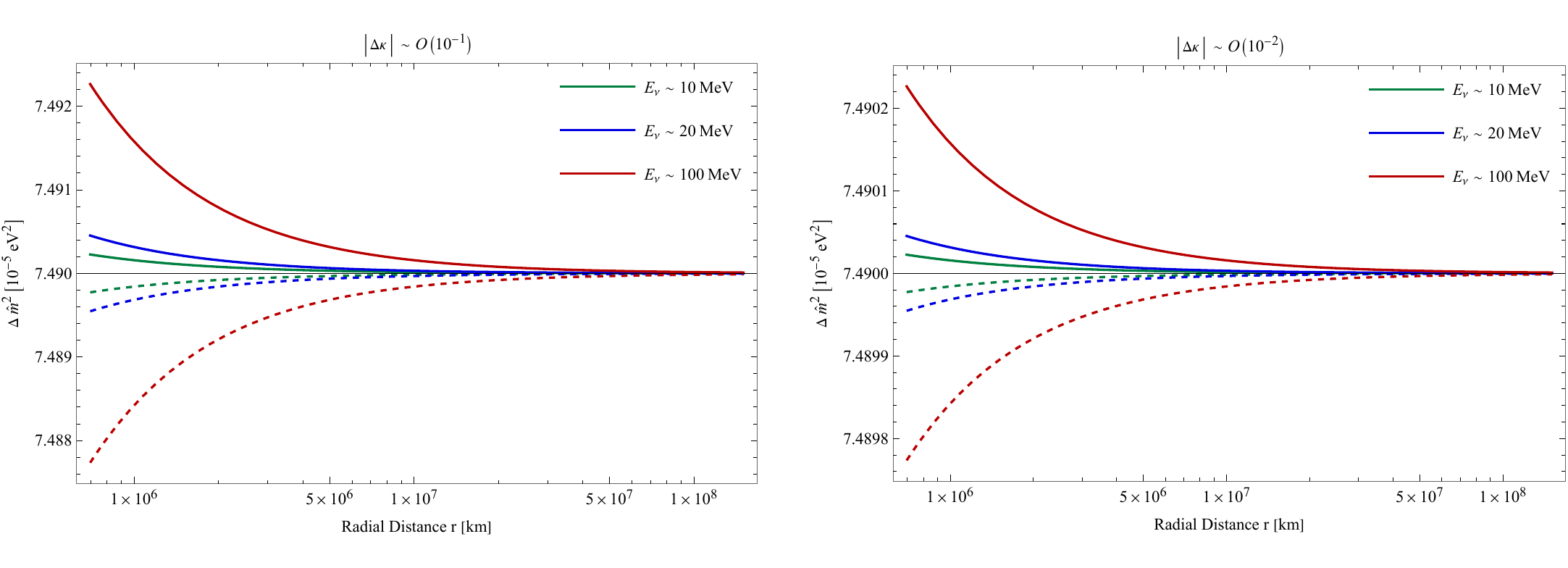}
	\caption{\footnotesize{Radial and energy dependence of the effective solar mass-squared splitting.
			The mass-squared splitting tends toward the solar surface for higher-energy neutrinos, converging to a constant value, corresponding to the LMA solution (cf. Table \ref{table1}).
			We have set $\alpha \sim\mathcal{O}(1~m^2)$ and two possible values for the coupling difference $|\Delta\kappa_{ij}| \sim\mathcal{O}(10^{-1})$ (left panel) and $|\Delta\kappa_{ij}| \sim\mathcal{O}(10^{-2})$ (right panel).
			Solid curves represent positive $\Delta\kappa_{ij}$ values, and dashed curves represent negative ones.
	}}
	\label{fig2}
\end{figure}

Furthermore, the torsion-neutrino coupling introduces an energy-dependent shift in $\Delta \hat{m}^2$ (see Eq.(\ref{eqn30})), with a deviation occurring at the solar surface for $^8$B neutrinos (with $E_\nu \simeq 10$ MeV).
It is important to note that this analysis does not include matter potential effects on $\Delta \hat{m}^2$.
This modification tends to the vacuum value on the Earth, i.e., $7.49 \times 10^{-5} \text{eV}^2$, consistent with large mixing angle (LMA) parameters.
The right panel of Figure \ref{fig2} (for $|\Delta\kappa_{ij}| \sim \mathcal{O}(10^{-2})$) reveals weaker but probably non-negligible corrections to the mass-squared splitting, suggesting detectability in high-precision future experiments.
As the plot indicates, a positive coupling difference $\Delta\kappa_{ij}>0$ (solid curves) results in a descending mass-squared splitting, while a negative coupling difference $\Delta\kappa_{ij}<0$ (dashed curves) produces an ascending trend.

\begin{figure}[H]
	\centering
	\includegraphics[scale=0.55]{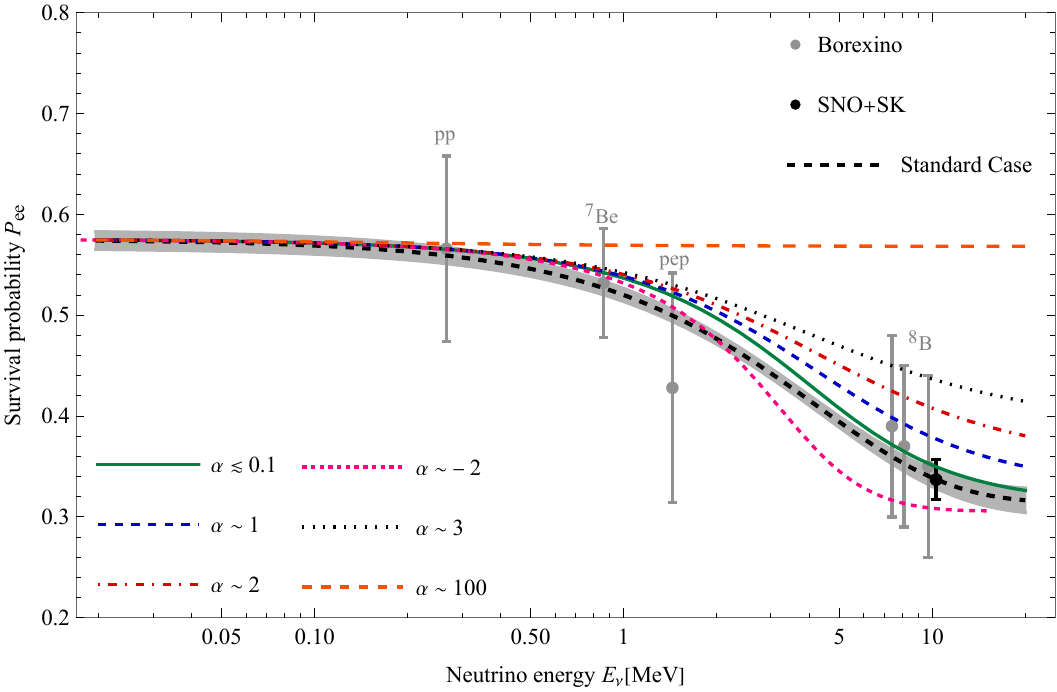}
	\caption{\footnotesize{Transition between matter and vacuum-dominated neutrino oscillation regimes.
			This figure is plotted for various values of $\alpha$ (in units $m^2$).
			We have assumed that $\Delta \kappa \simeq 0.001$.
			The observational data are taken from Borexino \cite{Borexino} (gray points) and SNO+SK \cite{SNO+SK} (a single black point).
			The black dashed curve represents the standard scenario, corresponding to the case where $(\alpha, \Delta\kappa) = (0,0)$.
	}}
	\label{fig3}
\end{figure}

\begin{figure}[H]
	\centering
	\includegraphics[scale=0.55]{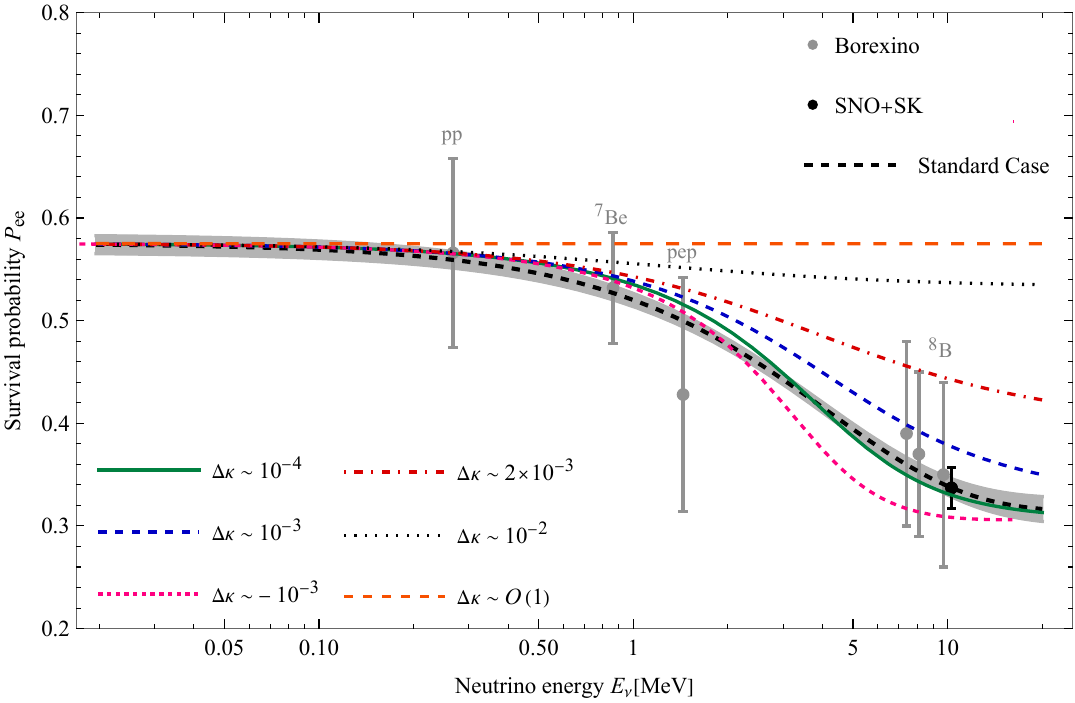}
	\caption{\footnotesize{This figure displays $P_{ee}$ versus neutrino energy $E_\nu$ for various selected $\Delta\kappa$ values and for $\alpha \sim \mathcal{O}(1~m^2)$.
	}}
	\label{fig4}
\end{figure}

We show the predictions of $P_{ee}(E_\nu)$ for solar neutrinos, spanning the energy regime from the lowest-energy $pp$ to the higher-energy $^8$B neutrinos.
Once passing through the dense region inside matter, the neutrino state is determined by the comparison between $\cos 2\theta$ and $\frac{2\hat{E}_\nu V(r_0)}{\Delta \hat{m}^2}$, evaluated at the neutrino's production point $r_0$.
The predicted survival probability can fall into three regimes:

$\blacksquare$ $\cos 2\theta \gg \frac{2\hat{E}_\nu V(r_0)}{\Delta \hat{m}^2}$:
For low-energy $pp$ neutrinos, where matter effects are negligible, neutrino propagation follows the vacuum oscillation pattern, yielding an averaged electron survival probability of $P_{ee} \simeq 0.55$.
In the present model, this vacuum-dominated regime may occur for $|\alpha| \gtrsim 100~m^2$ or $|\Delta\kappa| \gtrsim 0.1$, as shown in Figs. \ref{fig3} and \ref{fig4}, rendering the oscillation behavior of all solar neutrinos energy-independent.

However, comparison with Borexino data \cite{Borexino-Data} imposes constraints on the torsional parameter $\alpha$.
While enhanced matter effects significantly suppress the survival probability for higher-energy $^8$B neutrinos—effectively restoring vacuum oscillations at large $f(T)$ parameters ($|\alpha| \gtrsim 100~m^2$)—the model’s disagreement with observations restricts the model parameter to the range $-2\lesssim\alpha [m^2]\lesssim 3$.

$\blacksquare$ $\cos 2\theta \gtrsim \frac{2\hat{E}_\nu V(r_0)}{\Delta \hat{m}^2}$:
For this case, we encounter the neutrino mixing, modified by matter effects, which can be effectively modeled using adiabatic propagation.
Here, the survival probability for solar electron-neutrinos is given by
\begin{eqnarray}\label{eqn41}
P_{ee}(E_\nu) = \frac{1}{2} \left[1 + \cos2\theta \cos2\theta_m\right].
\end{eqnarray}
It is predicted that the torsion-induced effects might be considerable for this case, i.e., for $^8$B neutrinos with energy of around $\sim 10$ MeV, as they have intersections with the standard matter effects (represented by the gray bands with $1\sigma$ C.L. in Figs.{\ref{fig3}} and \ref{fig4}) and with the ``standard case'' $\alpha = 0$ (shown by the black dashed curve).
Similarly, the $\nu_e \to \nu_\mu$ transition probability during propagation is described by
\begin{eqnarray}\label{eqn42}
P_{e\mu}(E_\nu) = \frac{1}{2} \left[1 - \cos2\theta \cos2\theta_m\right].
\end{eqnarray}
Based on these, torsion-induced modifications offer a plausible explanation for the observed deficit of the solar $\nu_e$ flux.

$\blacksquare$ $\cos 2\theta < \frac{2\hat{E}_\nu V(r_0)}{\Delta \hat{m}^2}$:
For higher-energy neutrinos, they can cross the resonance region.
The $\nu_e$ survival probability is approximately $P_{ee} \simeq 0.33$ for this situation.

To quantify the best fitting in probability space, $\chi^2$ analysis is calculated using the formula
\begin{eqnarray}\label{eqn43}
\chi^2 = \sum_i \frac{\left(P_i^{\text{obs}} - P_i^{\text{th}}\right)^2}{\sigma_i^2}.
\end{eqnarray}
Here, $P_i^{\text{obs}}$ and $\sigma_i$ are the observed central values and uncertainties from the data (in Figs. \ref{fig3} and \ref{fig4}), and $P_i^{\text{th}}$ is the theoretically predicted value, obtained by the present model.
The best-fit values of the model parameters are determined by treating both $\alpha$ and $\Delta\kappa$ as free parameters and varying them simultaneously to locate the global minimum of the $\chi^2$ function against the experimental data.
The corresponding $1\sigma$ and $2\sigma$ confidence intervals are determined by the parameter regions where $\Delta\chi^2 = \chi^2 - \chi^2_{\text{bf}}$ does not exceed $1.0$ and $4.0$, respectively.
In this framework, all experimental uncertainties are treated as independent statistical errors; no correlated systematic effects across experiments or data releases are incorporated.
The analysis does not impose external theoretical priors on the model parameters $\alpha$ and $\Delta\kappa$; constraints arise exclusively from the solar neutrino data likelihood.
Table \ref{table2} summarizes constraints on the $f(T)$ gravity parameter $\alpha$ inferred from measurements of $^8$B neutrino flux by several generations of solar neutrino experiments.

The Kamiokande \cite{Kamiokande} result yields a best-fit value $\alpha_{\text{bf}} = 0.22$ with relatively broad $1\sigma$ and especially $2\sigma$ confidence intervals, implying its limited statistical precision compared to later experiments.
Therefore, Kamiokande provides only weak constraints on $\alpha$ and is fully consistent with $\alpha = 0$ limit.

A general observation is that experiments with higher statistical precision, notably the SK phases I-IV \cite{SK-I,SK-II,SK-III,Super-Kamiokande}, yield comparatively tight constraints on $\alpha$.
Individually, the SK phases favor positive best-fit values of $\alpha$ of order $\mathcal{O}(1~m^2)$, with best fits ranging from $\alpha_{\text{bf}} \simeq 0.55$ to $0.89$.
Importantly, all individual SK measurements are statistically consistent with $\alpha = 0$ at the $1\sigma$ level, as their confidence intervals include zero.
The global-SK (combined SK analysis) significantly sharpens these constraints, reducing the allowed parameter space to $[-0.03, 1.19]$ at $1\sigma$ and $[-0.62, 1.95]$ at $2\sigma$.

In contrast, SNO \cite{SNO-I,SNO-II,SNO-III} results display a marked dependence on the detection channel.
Charged-current (CC) measurements across all SNO phases consistently favor large negative best-fit values, $\alpha_{\text{bf}} \simeq -5.52$, which appear to saturate the lower boundary of the parameter space explored.
Their $1\sigma$ intervals lie entirely in the negative-$\alpha$ region, indicating tension with both $\alpha = 0$ and the SK-preferred values.
However, the corresponding $2\sigma$ intervals extend into positive $\alpha$ values, implying that this tension is not decisive at higher confidence levels.

By contrast, SNO measurements based on $\nu e$ scattering yield best-fit values of $\alpha$ close to zero or mildly positive and are broadly compatible with SK results within uncertainties.
The global-SNO fit (restricted to $\nu e$ phases) shows a positive best-fit value $\alpha_{\text{bf}} \simeq 4.55$ but is accompanied by very wide confidence intervals, reflecting the comparatively large experimental uncertainties of these channels.

The combined SK+SNO analysis yields a best-fit value $\alpha_{\text{bf}} \simeq 1.38$, with a $1\sigma$ interval that is entirely positive.
This result indicates that, when the dominant datasets are combined, the data exhibit a mild preference for positive $\alpha$, although $\alpha \to 0$ remains allowed at the $2\sigma$ level.

The Borexino \cite{Borexino,Borexino-Data} measurement, characterized by intermediate precision, provides weaker constraints: its best-fit value is negative $\alpha_{\text{bf}} \simeq -1.07$, but both its $1\sigma$ and $2\sigma$ ranges include zero, rendering it consistent with both standard teleparallel gravity (with $\alpha =0$) and modest deviations (with $\alpha \ne 0$).

Finally, the global-fit including all experiments yields a best-fit value $\alpha_{\text{bf}} \simeq 2.47$, with a $1\sigma$ interval $[1.09, 3.81]$ that excludes $\alpha=0$.
This suggests an overall preference for a positive, nonvanishing $f(T)$ gravity parameter when all available solar neutrino data are considered simultaneously.
Nevertheless, the $2\sigma$ interval $[-0.41, 5.14]$ still includes $\alpha = 0$, indicating that the evidence for deviations from the standard scenario remains suggestive rather than conclusive.
\begin{table}[H]
	\begin{center}
		\tiny
		\caption{\footnotesize{Results from solar neutrino experiments regarding $^8$B neutrino flux.
		Best-fit values and $1\sigma$ and $2\sigma$ confidence intervals for the model parameter $\alpha$ for the present model.
		}}
		\label{table2}
		\begin{tabular}{|c|c|c|c|c|}
			\hline
			\hline
			Experiment & $\nu_e$ flux [$10^6/\text{cm}^2 \text{s}$] & $\alpha [m^2]$ best-fit&  1$\sigma$ range $[m^2]$ & 2$\sigma$ range $[m^2]$\\
			\hline \hline
			Kamiokande \cite{Kamiokande} & $2.80 \pm 0.19$ &$+0.22$ & $[-0.56, +2.21]$ & $[-1.12, +45.7]$ \\ \hline
			SK-I \cite{SK-I} & $2.38 \pm 0.02$ &$+0.74$ & $[-0.49, +2.24]$ & $[-2.28, +4.73]$ \\
			SK-II \cite{SK-II} & $2.41 \pm 0.05$ &$+0.89$ & $[-0.36, +2.51]$ & $[-1.96, +5.33]$ \\
			SK-III \cite{SK-III} & $2.32 \pm 0.04$ &$+0.55$ & $[-0.73, +2.02]$ & $[-3.01, +4.26]$\\
			SK-IV \cite{Super-Kamiokande} & $+2.31 \pm 0.02$ &$+0.61$ & $[-0.69, +2.07]$ & $[-3.09, +4.24]$ \\
			Global-SK (SK I-IV) & &$+0.55$ & $[-0.03, +1.19]$ & $[-0.62,+1.95]$ \\ \hline
			SNO-Phase I (CC) \cite{SNO-I} & $1.76^{+0.06}_{-0.05}$ & $-5.53$ & $[-5.52, -1.19]$ & $[-5.52, +1.50]$ \\
			SNO-Phase I ($\nu e$) \cite{SNO-I} & $2.39^{+0.24}_{-0.23}$ & $+0.67$ & $[-0.94, +2.77]$ & $[-2.86, +7.64]$ \\
			SNO-Phase II (CC) \cite{SNO-II} & $1.68 \pm 0.06$ & $-5.52$ & $[-5.52,-1.93]$ & $[-5.52, +0.66]$ \\
			SNO-Phase II ($\nu e$) \cite{SNO-II} & $2.35 \pm 0.22$ & $+0.80$ & $[-1.01, +3.01]$ & $[-3.11, +7.59]$ \\
			SNO-Phase III (CC) \cite{SNO-III} & $1.67^{+0.05}_{-0.04}$ & $-5.52$ & $[-5.52, -2.10]$ & $[-5.52, +0.43]$ \\
			SNO-Phase III ($\nu e$) \cite{SNO-III} & $1.77^{+0.24}_{-0.21}$ & $-5.52$ & $[-5.52, +0.18]$ & $[-5.52, +4.10]$ \\
			Global-SNO (only $\nu e$ phases) &  & $+4.55$ & $[-2.17, +10.3]$ & $[-11.6, +16.7]$ \\ \hline
			SK+SNO &  & $+1.38$ & $[+0.55, +2.24]$ & $[-0.31, +3.19]$ \\ \hline
			Borexino  \cite{Borexino-Data,Borexino} & $2.57^{+0.17}_{-0.18}$ & $-1.07$ & $[-5.33, +1.86]$ & $[-9.80, +4.47]$ \\ \hline
			Global-fit & & $+2.47$ & $[+1.09, +3.81]$ & $[-0.41, +5.14]$ \\
			\hline
		\end{tabular}
	\end{center}
\end{table}
In summary, Table \ref{table2} demonstrates that higher-precision solar neutrino measurements, particularly from Super-Kamiokande, place stringent bounds on $\alpha$, while channel-dependent effects in SNO introduce some tension and broaden the global parameter space.
The combined analyses mildly favor positive values of $\alpha$, but the standard teleparallel gravity limit remains statistically viable 
at the $2\sigma$ level. 

Also, we may compare the constraints on $\alpha$ with findings from some gravitational experiments.
Stringent constraints on $\alpha$ are derived from observed secular variations in planetary orbits.
While a previous study \cite{Iorio} calculated a bound of $|\alpha| \lesssim 1.8 \times 10^4~m^2$ from the orbits of the inner solar system planets, current and more precise analyses of these orbital perturbations have significantly tightened this constraint to $|\alpha| \leq 1.2 \times 10^2~m^2$ \cite{SolSysTest1}.
A further refinement of this limit arises from independent orbital dynamics studies, with \cite{Ruggiero} reporting a significantly tighter constraint of $|\alpha| \leq 2.3 \times 10~m^2$.
The limits of $\alpha$ derived from our analysis are consistent with these reported constraints, as we have mostly obtained $|\alpha| \lesssim 5~m^2$, see Table \ref{table2}.

We also investigate the impact of modified matter effects on neutrino flavor transitions by computing the electron-neutrino survival probability across a range of coupling differences $\Delta\kappa$, see Figure \ref{fig4}.
This analysis explores how the coupling strength parameter $\Delta\kappa$ modifies neutrino flavor transitions with four regimes:

$\bullet$ For $0 < \Delta\kappa < 10^{-4}$, the curves imply standard MSW behavior with minimal deviations.

$\bullet$ The transitional regime ($10^{-4}\lesssim\Delta\kappa \lesssim 2 \times 10^{-3}$) shows nonlinear effects like shifted resonances, offering optimal sensitivity to the present model.

$\bullet$ The system exhibits oscillations in vacuum as a result of strong coupling differences, typically defined by values of $\Delta\kappa \gtrsim \mathcal{O}(0.1)$.

$\bullet$ Finally, negative coupling differences ($\Delta\kappa < 0$) lead to a suppression of $P_{ee}$ (pink-dashed curve),
again indicating the dominance of the new neutrino-torsion coupling.

These effects stem from $\Delta\kappa$'s role in scaling the matter potential, with positive/negative values enhancing or suppression conventional oscillations.
Current solar neutrino data constrain $|\Delta\kappa| \lesssim 2 \times 10^{-3}$, while future experiments like JUNO \cite{JUNO}
may impose tighter constraints.

Table \ref{table3} presents constraints on the neutrino-torsion coupling difference $\Delta\kappa$, as inferred from flux measurements by a variety of solar neutrino experiments.
For each dataset, as in the previous case, the table reports the best-fit value of $\Delta\kappa$ together with its corresponding $1\sigma$ and $2\sigma$ confidence intervals within the adopted theoretical framework.

The earliest Kamiokande measurement yields a comparatively large best-fit value, $\Delta\kappa_{\text{bf}} \simeq 4.9 \times 10^{-3}$, and provides only a lower bound on the parameter.
The absence of an upper constraint and the relatively high best-fit value reflect the limited statistical precision of this experiment, implying that Kamiokande alone has weak discriminatory power and indicates that sizable values of $\Delta\kappa$ cannot be excluded.

The SK phases I-IV exhibit remarkable internal consistency.
Individually, all SK phases favor positive best-fit values of $\Delta\kappa_{\text{bf}}$ around $ (1.5-1.7)\times 10^{-3} $, with overlapping confidence intervals.
While the $2\sigma$ ranges of the individual phases extend into negative values, their $1\sigma$ intervals remain strictly positive, suggesting a mild but systematic preference for nonzero $\Delta\kappa$.
This preference is significantly reinforced in the combined SK analysis, which yields $\Delta\kappa = (1.69^{+0.58}_{-0.53})\times 10^{-3}$ at $1\sigma$.
Notably, the global-SK $2\sigma$ interval remains entirely positive, effectively implying $\Delta\kappa \ne 0$ within this dataset.

Results from SNO again display a strong dependence on the detection channel.
The CC measurements across all SNO phases converge to a common best-fit value, $\Delta\kappa_{\text{bf}} \simeq 6.2 \times 10^{-4}$, with relatively narrow upper bounds but confidence intervals that include $\Delta\kappa = 0$ at the $1\sigma$ or $2\sigma$ level.

In contrast, the SNO $\nu e$ scattering prefer larger and positive best-fit values comparable to those from SK, albeit with substantially broader uncertainties.
When all SNO data are combined, the resulting global-SNO constraint yields a small positive best-fit value, $\Delta\kappa_{\text{bf}} \simeq 2.5 \times 10^{-4}$, which remains compatible with zero within $1\sigma$, indicating that SNO alone does not provide decisive evidence for a nonvanishing coupling difference.

The combined SK+SNO analysis leads to a best-fit value $\Delta\kappa_{\text{bf}} \simeq 9.8 \times 10^{-4}$, with relatively tight confidence intervals that are strictly positive even at the $2\sigma$ level.
This result reflects the dominant statistical weight of the high-precision SK data while incorporating the complementary information from SNO, and it strengthens the indication of a positive neutrino-torsion coupling difference.

The Borexino measurement, characterized by larger experimental uncertainties, yields a small best-fit value and extremely wide confidence intervals that span both positive and negative values over a broad range.
Consequently, Borexino places only weak constraints on $\Delta\kappa$ and remains fully consistent with both the null hypothesis and the values favored by SK and SK+SNO.

Finally, the global-fit including all experiments yields a positive best-fit value $\Delta\kappa_{\text{bf}} \simeq 4.1 \times 10^{-4}$, with a $1\sigma$ interval that excludes zero $\Delta\kappa$.
Even at the $2\sigma$ level, the allowed range remains strictly positive.
This global result therefore provides overall evidence, albeit moderate, for a nonzero neutrino-torsion coupling difference, primarily driven by the high-precision SK data and their consistency across multiple phases.
\begin{table}[H]
	\begin{center}
		\tiny
		\caption{\footnotesize{Constraints on neutrino-torsion coupling difference $\Delta\kappa$ (in units of $10^{-4}$), showing best-fit values with 1$\sigma$ and 2$\sigma$ confidence intervals.}}
		\label{table3}
		\begin{tabular}{|c|c|c|c|c|}
			\hline
			\hline
			Experiment & $\nu_e$ flux [$10^6/\text{cm}^2 \text{s}$] & $\Delta \kappa$ best-fit $[10^{-4}]$&  1$\sigma$ range $[10^{-4}]$& 2$\sigma$ range $[10^{-4}]$\\
			\hline \hline
			Kamiokande \cite{Kamiokande} & $2.80 \pm 0.19$ & $48.95$ & $> 24.08$ & $> 6.300$ \\ \hline
			SK-I \cite{SK-I} & $2.38 \pm 0.02$ & $16.33$ & $[6.620,28.28]$ & $[-7.540,47.92]$ \\
			SK-II \cite{SK-II} & $2.41 \pm 0.05$ & $16.49$ & $[7.160,28.48]$ & $[-4.680,49.46]$ \\
			SK-III \cite{SK-III} & $2.32 \pm 0.04$ & $15.46$ & $[4.880,27.52]$ & $[-13.86,45.90]$\\
			SK-IV \cite{Super-Kamiokande} & $2.31 \pm 0.02$ & $14.77$ & $[4.900,25.92]$ & $[-13.42,42.44]$ \\
			Global-SK (SK I-IV) &  & $16.94$ & $[11.64,22.74]$ & $[6.380,29.52]$ \\ \hline
			SNO-Phase I (CC) \cite{SNO-I} & $1.76^{+0.06}_{-0.05}$ & $6.217$ & $[-0.220,6.200]$ & $[-4.200,6.200]$ \\
			SNO-Phase I ($\nu e$) \cite{SNO-I} & $2.39^{+0.24}_{-0.23}$ & $16.97$ & $[3.280,34.84]$ & $[-13.06,76.22]$ \\
			SNO-Phase II (CC) \cite{SNO-II} & $1.68 \pm 0.06$ & $6.217$ & $[0.900,6.200]$ & $[-2.940,6.200]$ \\
			SNO-Phase II ($\nu e$) \cite{SNO-II} & $2.35 \pm 0.22$ & $14.67$ & $[2.140,29.98]$ & $[-12.28,61.64]$ \\
			SNO-Phase III (CC) \cite{SNO-III} & $1.67^{+0.05}_{-0.04}$ & $6.217$ & $[1.140,6.200]$ & $[-2.600,6.200]$ \\
			SNO-Phase III ($\nu e$) \cite{SNO-III} & $1.77^{+0.24}_{-0.21}$ & $6.217$ & $[-2.240,6.200]$ & $[-8.040,6.200]$ \\
			Global-SNO &  & $2.534$ & $[-0.360,5.020]$ & $[-4.420,7.780]$ \\ \hline
			SK+SNO &  & $9.781$ & $[6.780,12.90]$ & $[3.700,16.34]$ \\ \hline
			Borexino  \cite{Borexino-Data,Borexino} & $2.57^{+0.17}_{-0.18}$ & $2.996$ & $[-46.64,37.24]$ & $[-98.50,67.50]$ \\ \hline
			Global-fit & & $4.082$ & $[2.600,5.500]$ & $[0.9800,6.940]$ \\
			\hline
		\end{tabular}
	\end{center}
\end{table}

\begin{figure}[H]
	\begin{subfigure}{.5\textwidth}
		\centering
		\includegraphics[scale=0.35]{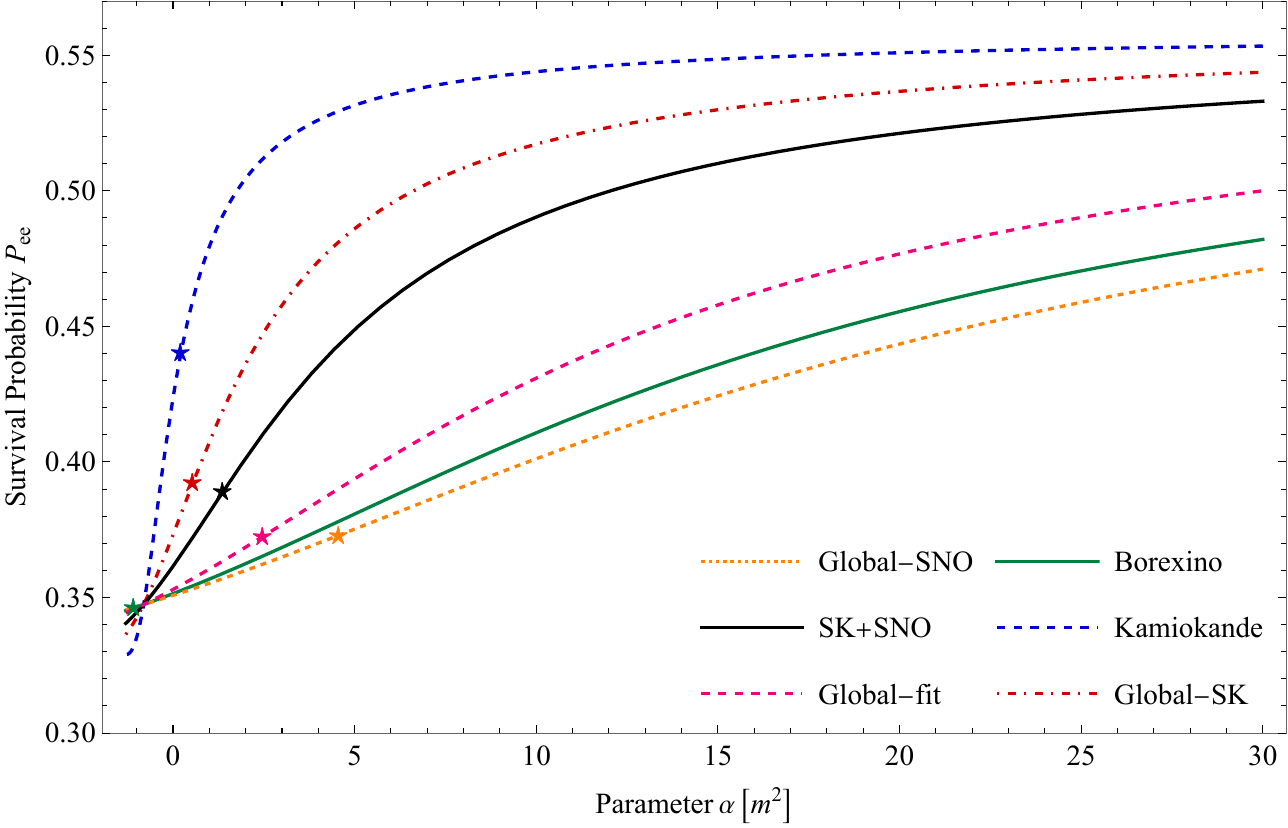}
		\label{fig5-1}
		\caption{}
	\end{subfigure}
	\begin{subfigure}{.5\textwidth}
		\centering
		\includegraphics[scale=0.35]{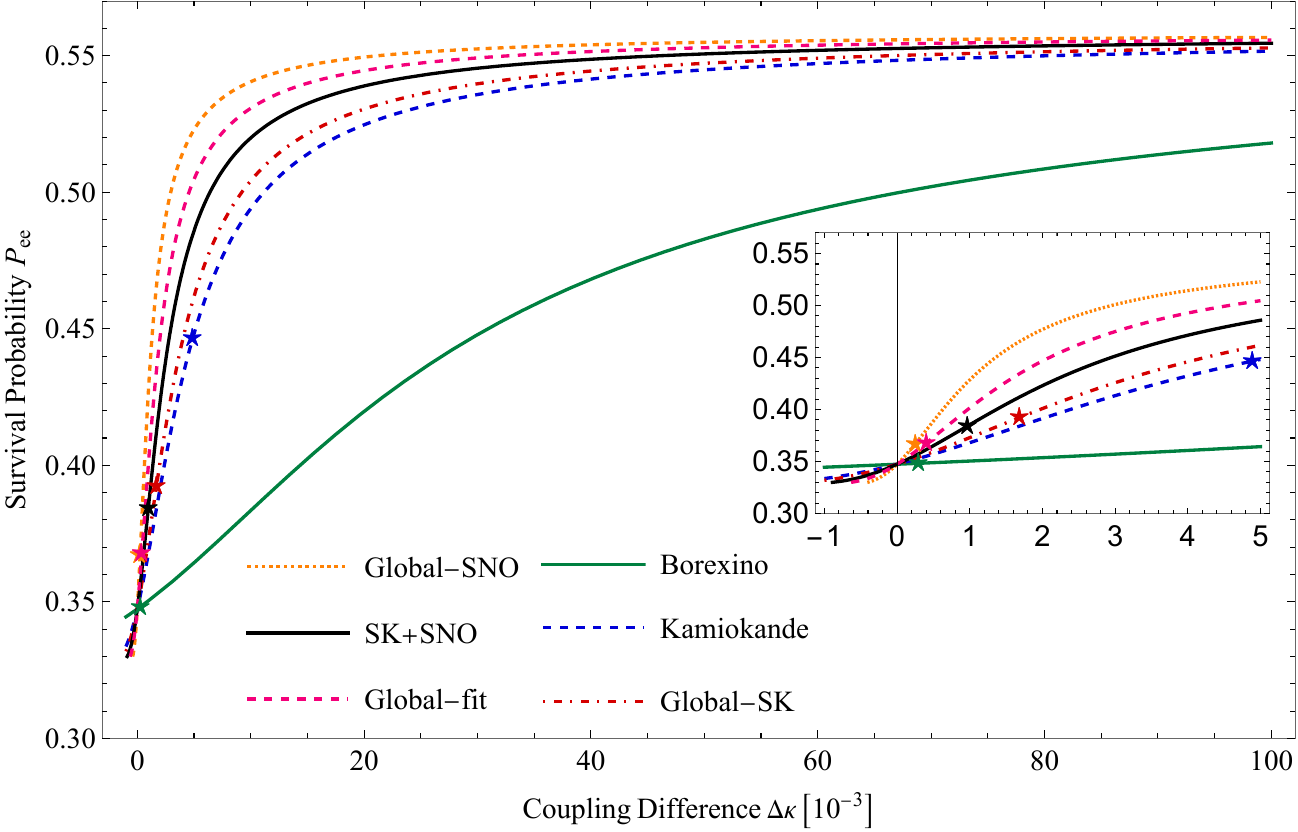}
		\label{fig5-2}
		\caption{}
	\end{subfigure}
	\caption{\footnotesize{(a) Matter-to-vacuum transition in neutrino oscillations under $f(T)$ gravity.
			The plot displays the evolution of the survival probability $P_{ee}$ across varying strengths of the model parameter $\alpha[m^2]$, for each observational dataset.
(b) $P_{ee}$ as a function of the coupling difference $\Delta\kappa [10^{-3}]$.
Best-fit values are shown by the asterisks in both plots.
}}
	\label{fig5}
\end{figure}

The explicit dependence of $P_{ee}$ on the parameters $\alpha$ and $\Delta\kappa$ is illustrated over the complete parameter space for all experiments listed in Tables \ref{table2} and \ref{table3}.
Figure \ref{fig5}(a) illustrates the dependence of the electron-neutrino survival probability $P_{ee}$ on the parameter $\alpha$ for $10$ MeV $^8$B neutrinos.
As $\alpha$ increases, matter-induced effects encoded in the $f(T)$ framework become progressively suppressed, leading to a smooth rise in $P_{ee}$.
In the regime $\alpha \gg 1~m^2$, the survival probability saturates at $P_{ee}\simeq 0.55$, indicating the recovery of the vacuum oscillation limit.
The agreement with global solar neutrino data further supports the physical viability of the chosen parameter range.
Figure \ref{fig5}(b) presents the dependence of the electron-neutrino survival probability $P_{ee}$ on the coupling difference $\Delta\kappa$.
For very small coupling differences $\Delta\kappa \ll 1$, the survival probability attains its minimum, indicating a regime dominated by standard matter effects.
As $\Delta\kappa$ increases, deviations induced by the modified coupling become significant, leading to a systematic enhancement of $P_{ee}$.
This trend reflects the growing influence of profile-dependent corrections in the neutrino propagation dynamics.

\section{Conclusion}\label{Sec5}
This work has established a theoretical framework within $f(T)$ gravity (see Section \ref{Sec2}) to quantify the impact of spacetime torsion on neutrino flavor change.
We have derived general, analytical expressions that describe how torsion induces a measurable phase shift and neutrino wavefunction, see Section \ref{Sec3}.
Crucially, these modifications affect both the vacuum oscillation paradigm (Subsection \ref{Subsec3.1}) and the standard MSW resonance mechanism in matter (Subsection \ref{Subsec3.2}).
Within the proposed framework, applied to the polynomial density profile, the modification of the neutrino flavor deficit due to torsion provides a mechanism to address potential observational deficit in flavor conversion, without invoking new physics or NSI.
Furthermore, the torsion-induced modification to the effective mass-squared splitting introduces an environmental dependence governed by the local torsional background.
Such a dependence offers a plausible explanation for the apparent variation between oscillation parameters extracted from solar and terrestrial neutrino measurements.

Moreover, the formalism provides a direct link between the geometric formulation of gravity and quantum particle phenomenology.
This paper demonstrates a method for investigating theories beyond teleparallel equivalent of general relativity.
It establishes calculations of neutrino behavior in astrophysical environments as a possibility for testing higher-order terms of torsion in $f(T)$ gravity.
The analysis places stringent limits on the key parameter $\alpha$ of this model, finding mostly $|\alpha| \lesssim 5~m^2$
(2$\sigma$ allowed ranges from the numerical results are given in Table \ref{table2}). 

Based on discussion after the Table \ref{table2}, our constraints on $\alpha$ from solar neutrinos are stronger than those from planetary orbits, reflecting the high sensitivity of quantum interferometric phenomena to small geometric perturbations over astronomical distances.
This result is consistent with constraints from other studies \cite{Ruggiero,Xie,Iorio}, which found $|\alpha| \lesssim 23~m^2$ for the most stringent one. 

Future work will involve applying these constraints to specific astrophysical objects and refining the predictions for next-generation neutrino observatories.

\section*{Acknowledgement}
This work is based upon research funded by Iran National Science Foundation (INSF) under project No. 4036948.

\end{document}